# The Parikh functions of sparse context-free languages are quasi-polynomials *

Flavio D'Alessandro
Dipartimento di Matematica,
Università di Roma "La Sapienza"
Piazzale Aldo Moro 2, 00185 Roma, Italy.
e-mail: dalessan@mat.uniroma1.it,
http://www.mat.uniroma1.it/people/dalessandro

Benedetto Intrigila
Dipartimento di Matematica,
Università di Roma "Tor Vergata",
via della Ricerca Scientifica, 00133 Roma, Italy.
e-mail intrigil@mat.uniroma2.it,

Stefano Varricchio
Dipartimento di Matematica,
Università di Roma "Tor Vergata",
via della Ricerca Scientifica, 00133 Roma, Italy.
e-mail varricch@mat.uniroma2.it,
http://www.mat.uniroma2.it/~varricch

**Abstract**

Let $L$ be a sparse context-free language over a finite alphabet $A = \{a_1, \ldots, a_t\}$ and let $f_L : \mathbb{N}^t \to \mathbb{N}$ be its Parikh (counting) function. We prove that the map $f_L$ is a box spline, that is there exists a finite set $\Pi$ of hyperplanes of $\mathbb{R}^t$, through the origin, such that $f_L$ is a quasi-polynomial on every polyhedral cone determined by $\Pi$. Moreover we prove that the Parikh function of such a language is rational.

*This work was partially supported by MIUR project ''Linguaggi formali e automi: teoria e applicazioni''. The first author acknowledges the partial support of fundings ''Facoltà di Scienze MM. FF. NN. 2006'' of the University of Rome ''La Sapienza'' and fundings CRUI DAAD Program ''Vigoni 2006''.

# 1 Introduction

Given a word $w$ over an alphabet $A = \{a_1, \ldots, a_t\}$, *the Parikh vector* of $w$ is the vector $(n_1, \ldots, n_t)$, where for every $i$, with $1 \leq i \leq t$, $n_i$ is the number of occurrences of letter $a_i$ in $w$. Given a language $L$ over $A$, the *Parikh map* $f_L$ is the map which associates with non negative integers $n_1, \ldots, n_t$, the number $f_L(n_1, \ldots, n_t)$ of all the words in $L$ having Parikh vector equal to $(n_1, \ldots, n_t)$. If we impose restrictions on the growth rate of $f_L$ we obtain different classes of languages. In particular, a language is termed *Parikh slender* (see [13]) if there is a positive integer $r$ such that for every $n_1, \ldots, n_t$, $f_L(n_1, \ldots, n_t) \leq r$ holds.

In this paper we shall consider the Parikh map of *bounded context-free languages.* Bounded context-free languages and their properties have been extensively studied in [9], where, in particular, one proves that it is decidable whether a given context-free language is bounded. It is also known that bounded context-free languages are exactly the *sparse* context-free languages [21]. A formal language $L$ is termed *sparse* if its counting function is upper-bounded by a polynomial or, equivalently, if its Parikh function is upper bounded by a multivariate polynomial. It is clear that the class of sparse context-free languages include that of Parikh slender ones.

A *quasi-polynomial* is a a map $F : \mathbb{N}^t \to \mathbb{N}$ defined by a finite family of multivariate polynomials, with rational coefficients, $\{p_{(d_1,d_2,\cdots,d_t)} \mid d_1, \ldots, d_t \in \mathbb{N},\ 0 \leq d_i < d\}$, where, for every $(x_1, \ldots, x_t) \in \mathbb{N}$, if $d_i$ is the reminder of the division of $x_i$ by $d$, one has:

$$F(x_1, \ldots, x_t) = p_{(d_1,d_2,\cdots,d_t)}(x_1, \ldots, x_t).$$

In this paper we prove that the Parikh map $f_L$ of a sparse context free language $L$ can be exactly calculated using a finite number of quasi-polynomials. More precisely, if $L$ is a sparse context-free language, then there exist a partition of $\mathbb{N}^t$ into a finite number of polyhedral conic regions $R_1, \ldots, R_s$, determined by hyperplanes, through the origin, with rational equations, and a finite number of quasi-polynomials $p_1, \ldots p_s$, such that for any $(n_1, \ldots, n_t)$ one has:
$f_L(n_1, \ldots, n_t) = p_j(n_1, \ldots, n_t)$ where $j$ is such that $(n_1, \ldots, n_t) \in R_j$.

This is obtained by reducing the computation of the Parikh map of $L$ to the computation of the number of non-negative solutions of a system of diophantine linear equations of the form

$$\sum_{j=1}^{n} a_{ij} x_j = n_i, \quad 1 \leq i \leq t,\ a_{ij} \in \mathbb{N}, \tag{1}$$

as a function $F(n_1, \ldots, n_t)$ of the constants $n_1, \ldots, n_t$.
The latter computation deserves a special mention. It was first considered in the context of Numerical Analysis where, in a celebrated paper by Dahmen and Micchelli [3], it has been proved that the counting function of a Diophantine system of linear equations can be described by a set of quasi-polynomials, under suitable conditions on the matrix of the system. Recently this result has been object of further investigations in [4, 5, 24] where important theorems on the algebraic and combinatorial structure of partition functions have been obtained.

In this paper, we present a combinatorial proof of the description of the counting map of the system (1). This proof, which appears to be new, is of elementary character, effective and makes the paper self-contained completely. We remark that the regions $R_1, \ldots, R_s$ as well as the quasi-polynomials $p_1, \ldots, p_s$ that gives a description of the Parikh map can be effectively computed from an effective presentation of the language $L$. The decidability of some problems on the Parikh map of context-free languages is an easy consequence of our main result. In particular, one can decide whether a context-free language is Parikh thin or Parikh slender [13].

In the proof of our main result, some important and deep theorems of Formal Language Theory have been used. More precisely, the combinatorial method developped to describe the Parikh map of a context-free language is based upon the representation of such languages as bounded semi-linear sets. In particular, two important results have been used: the celebrated Cross-Section Theorem by Eilenberg and the theorem, by Eilenberg and Schützenberger, which characterizes the rational subsets of $\mathbb{N}^t$ (*cf* [7, 23]). These theorems provide a crucial tool in order to cope with the ambiguity of context-free languages. In this context, we also recall another important recent result that gives a characterization of sparse context-free languages in terms of finite unions of *Dyck loops* (*cf* [16, 17]). However, this latter result cannot be used to compute any counting function because of the ambiguity of the representation of such a language as a finite union of Dyck loops. Indeed, consider the language $L = \{a^n b^m c^m d^n \mid n, m \geq 0\} \cup \{a^n b^n c^m d^m \mid n, m \geq 0\}$. The language $L$ is sparse context-free but it cannot be represented unambiguously as a finite union of Dyck loops.

We complete our analysis proving that the Parikh function of a sparse context-free language is rational. This result is remarkable since, as it is well known [8], the Parikh function of a context-free language may be transcendent.

## 2 On systems of diophantine equations.

The aim of this section is to present some results that we will use as our main tools in the sequel of the paper. We assume that the reader is familiar with the basic notions of rational, context-free and semi-linear languages. The reader is referred to [1, 6, 9, 14, 23]. We start this section by proving an important theorem concerning systems of diophantine equations. For this purpose, we recall some preliminary definitions and results.

**Lemma 1** *Let $q(x_1, \ldots, x_t, x)$ be a polynomial in $t+1$ variables with rational coefficients and let*

$$F : \mathbb{N}^t \times \{\{-1\} \cup \mathbb{N}\} \longrightarrow \mathbb{Q}$$

*be the map defined as:*

$$F(x_1,\dots,x_t,x) = \begin{cases} \sum_{\lambda=0,\dots,x} q(x_1,\dots,x_t,\lambda) & x \geq 0, \\ 0 & x = -1. \end{cases}$$

*There exists a polynomial* $p(x_1,\dots,x_t,x)$ *in* $t+1$ *variables with rational coefficients such that, for every* $(x_1,\dots,x_t,x) \in \mathbb{N}^t \times \{\{-1\} \cup \mathbb{N}\}$*, one has:*

$$F(x_1,\dots,x_t,x) = p(x_1,\dots,x_t,x).$$

*Proof.* Write $q(x_1,\dots,x_t,x)$ as:

$$a_0 + a_1 x + \cdots + a_n x^n, \tag{2}$$

where, for every $i = 0,\dots,n$, $a_i$ is a suitable polynomial in the variables $x_1,\dots,x_t$ with rational coefficients. By Eq. (2), if $x \geq 0$, for every $x_1,\dots,x_t \in \mathbb{N}$, one has:

$$F(x_1,\dots,x_t,x) = \sum_{\lambda=0,\dots,x} q(x_1,\dots,x_t,\lambda) = \sum_{j=0,\dots,n} \left( a_j \cdot \sum_{\lambda=0,\dots,x} \lambda^j \right). \tag{3}$$

On the other hand, by using a standard argument (*cf* Lemma 15 of the Appendix), one can prove that, for any $j \in \mathbb{N}$, there exists a polynomial $p_j(x)$ with rational coefficients in one variable $x$ such that:

3.1) for any $x \in \mathbb{N}$, $p_j(x) = \sum_{\lambda=0,\dots,x} \lambda^j$.

3.2) $p_j(-1) = 0$.

For any $j = 0,\dots,n$, let $p_j$ be the polynomial defined above and let $p = p(x_1,\dots,x_t,x)$ be the polynomial defined as:

$$p = \sum_{j=0,\dots,n} a_j p_j.$$

Then by Eq. (3.2), one has $p(x_1,\dots,x_t,-1) = 0$. Moreover, for every $x \geq 0$, by Eq. (3) and (3.1), one has:

$$F(x_1,\dots,x_t,x) = \sum_{j=0,\dots,n} \left( a_j \cdot \sum_{\lambda=0,\dots,x} \lambda^j \right) = \sum_{j=0,\dots,n} a_j p_j(x) = p(x_1,\dots,x_t,x).$$

The proof is thus complete. □

**Definition 1** *A map* $F : \mathbb{N}^t \longrightarrow \mathbb{N}$ *is said to be a* quasi-polynomial *if there exist* $d \in \mathbb{N}$*,* $d \geq 1$*, and a family of polynomials in* $t$ *variables with rational coefficients:*

$$\{p_{(d_1,d_2,\cdots,d_t)} \mid d_1,\dots,d_t \in \mathbb{N},\ 0 \leq d_i < d\},$$

*where, for every* $(x_1, \ldots, x_t) \in \mathbb{N}^t$*, if* $d_i$ *is the reminder of the division of* $x_i$ *by* $d$*, one has:*

$$F(x_1, \ldots, x_t) = p_{(d_1, d_2, \cdots, d_t)}(x_1, \ldots, x_t).$$

*The number* $d$ *is called the* period *of* $F$*.*

To simplify the notation, the polynomial $p_{(d_1, d_2, \cdots, d_t)}$ is denoted $p_{d_1 d_2 \cdots d_t}$.

**Definition 2** *Let* $F : \mathbb{N}^t \longrightarrow \mathbb{N}$ *be a map. Given a subset* $C$ *of* $\mathbb{N}^t$*,* $F$ *is said to be a* quasi-polynomial over $C$ *if there exists a quasi-polynomial* $q$*, such that* $F(x) = q(x)$*, for any* $x \in C$*.*

**Lemma 2** *The sum of a finite family of quasi-polynomials is a quasi-polynomial.*

*Proof.* It suffices to prove the claim for two quasi-polynomials. Let $f_1, f_2 : \mathbb{N}^t \longrightarrow \mathbb{N}$ be quasi-polynomials of periods $d_1, d_2$ respectively and let

$$\{p_{a_1 \cdots a_t} \mid \forall\, i = 0, \ldots, t,\ 0 \le a_i \le d_1 - 1\}, \text{ and}$$

$$\{q_{b_1 \cdots b_t} \mid \forall\, i = 0, \ldots, t,\ 0 \le b_i \le d_2 - 1\}$$

be the families of polynomials that define $f_1$ and $f_2$ respectively. Define a new quasi-polynomial $f$ as follows. Take $d = d_1 d_2$ as the period of $f$ and, for every $(c_1, \ldots, c_t) \in \{0, 1, \ldots, d-1\}^t$, take

$$f_{c_1 \cdots c_t} = p_{a_1 \cdots a_t} + q_{b_1 \cdots b_t},$$

where, for any $i = 1, \ldots, t$, $a_i$ and $b_i$ are the reminders of the division of $c_i$ by $d_1$ and $d_2$ respectively. It is easily checked that the quasi-polynomial $f$ is the sum of $f_1$ and $f_2$. Indeed, if $x = (x_1, \ldots, x_t) \in \mathbb{N}^t$ and, for every $i = 1, \ldots, t$, $x_i \equiv c_i \bmod d$, then one has

$$c_i \equiv a_i \bmod d_1 \iff x_i \equiv a_i \bmod d_1$$

$$c_i \equiv b_i \bmod d_2 \iff x_i \equiv b_i \bmod d_2.$$

Therefore, if $x = (x_1, \ldots, x_t) \in \mathbb{N}^t$ and $x_i \equiv c_i \bmod d$, then we have:

$$f(x) = f_{c_1 \cdots c_t}(x) = p_{a_1 \cdots a_t}(x) + q_{b_1 \cdots b_t}(x) = f_1(x) + f_2(x).$$

The claim is thus proved. □

**Lemma 3** *Let* $F : \mathbb{N}^t \longrightarrow \mathbb{N}$ *be a map,* $d$ *be a positive integer, and* $C$ *be a subset of* $\mathbb{N}^t$*. If there exists a family of quasi-polynomials* $\{F_{d_1 d_2 \cdots d_t} \mid d_1, \ldots, d_t \in \mathbb{N},\ 0 \le d_i < d\}$*, such that , for every* $(x_1, \ldots, x_t) \in C$*, with* $x_i \equiv d_i \bmod d$*, one has:* $F(x_1, \ldots, x_t) = F_{d_1 d_2 \cdots d_t}(x_1, \ldots, x_t)$*, then* $F$ *is a quasi-polynomial over* $C$*.*

*Proof.* Let $k$ be the least common multiple of $d$ and of the periods of the quasi-polynomials of the set $\{F_{d_1 d_2 \cdots d_t} \mid d_1, \ldots, d_t \in \mathbb{N},\ 0 \leq d_i < d\}$. Let $(r_1, \ldots, r_t)$ be a tuple of $\{0, 1, \ldots k-1\}^t$ and let $(x_1, \ldots, x_t) \in C$ be such that, for every $i = 1, ..., t$, $x_i \equiv r_i \bmod k$. Then one can check that $F(x_1, \ldots, x_t) = q(x_1, \ldots, x_t)$ where $q$ is a polynomial uniquely determined by $(r_1, \ldots, r_t)$. Indeed, one can first observe that the tuple $(r_1, \ldots, r_t)$ uniquely determines, for every $i = 1, ..., t$, the reminder $d_i$ of the division of $x_i$ by $d$ since $d_i \equiv r_i \bmod d$. By hypothesis, one has $F(x_1, \ldots, x_t) = F_{d_1 d_2 \cdots d_t}(x_1, \ldots, x_t)$. Since $k$ is a multiple of the period of the quasi-polynomial $F_{d_1 d_2 \cdots d_t}$, the tuple $(r_1, \ldots, r_t)$ also determines a polynomial $q$ in the family of polynomials associated with $F_{d_1 d_2 \cdots d_t}$, such that $F(x_1, \ldots, x_t) = F_{d_1 d_2 \cdots d_t}(x_1, \ldots, x_t) = q(x_1, ..., x_t)$. The proof is thus complete. □

**Lemma 4** *Let $\lambda : \mathbb{N}^t \longrightarrow \mathbb{Q}$ be a map such that, for any $(x_1, \ldots, x_t) \in \mathbb{N}^t$,*

$$\lambda(x_1, \ldots, x_t) = b_1 x_1 + \cdots + b_t x_t,$$

*where $b_1, \ldots, b_t$ are given rational coefficients. Let $C, C'$ be subsets of $\mathbb{N}^t$ and let $a_1, \ldots, a_t$ be non negative integers such that the following properties are satisfied: for any $(x_1, \ldots, x_t) \in C$, one has*

- $\lambda(x_1, \ldots, x_t) \geq 0$,
- *if $\lambda \in \mathbb{N}$ and $\lambda < \lambda(x_1, \ldots, x_t)$, then $(x_1 - \lambda a_1, x_2 - \lambda a_2, \ldots, x_t - \lambda a_t) \in C'$.*

*Let $p$ be a quasi-polynomial over $C'$ and define the map $F$ as:*

$$F(x_1, \ldots, x_t) = \sum_{0 \,\leq\, \lambda \,<\, \lambda(x_1, \ldots, x_t)} p(x_1 - \lambda a_1, x_2 - \lambda a_2, \ldots, x_t - \lambda a_t).$$

*Then $F$ is a quasi-polynomial over $C$.*

*Proof.* Let $d \geq 1$ be the period of the quasi-polynomial $p$. Let $p_{d_1 d_2 \cdots d_t}$, with $0 \leq d_i \leq d-1$, be the polynomials defining $p$. Consider the set of integers $\mu$:

$$0 \leq \mu \ \leq \ \lceil \lambda(x_1, \ldots, x_t) \rceil - 1,$$

and consider on it the partition:

$$F_0(x_1, \ldots, x_t) \cup F_1(x_1, \ldots, x_t) \cup \cdots \cup F_{d-1}(x_1, \ldots, x_t), \tag{4}$$

defined as: for any $j = 0, \ldots, d-1$:

$$\mu \in F_j(x_1, \ldots, x_t) \quad \Longleftrightarrow \quad 0 \leq \mu \ \leq \ \lceil \lambda(x_1, \ldots, x_t) \rceil - 1 \text{ and } \mu \ \equiv \ j \ (\text{mod } d).$$

By Eq. (4), for every $(x_1, \ldots, x_t) \in C$, we have:

$$F(x_1, \ldots, x_t) \ = \ \sum_{j=0}^{d-1} S_j(x_1, \ldots, x_t), \tag{5}$$

where, for any $j = 0, \dots, d-1$:

$$S_j(x_1, \dots, x_t) = \sum_{\mu \in F_j(x_1,\dots,x_t)} p(x_1 - \mu a_1, x_2 - \mu a_2, \dots, x_t - \mu a_t).$$

Now we prove that, for any $j = 0, \dots, d-1$, $S_j$ is a quasi-polynomial over $C$.

Let us fix a tuple $(d_1, \dots, d_t) \in \{0, 1, \dots d-1\}^t$. Let $(x_1, \dots, x_t)$ be such that $x_i \equiv d_i \bmod d$. Then for any $\mu \in F_j(x_1, \dots, x_t)$ one has $x_i - \mu a_i \equiv d_i - j a_i \bmod d$. Now set $q = p_{c_1 c_2 \cdots c_t}$, where $(c_1, \dots, c_t) \in \{0, 1, \dots d-1\}^t$ and $c_i \equiv d_i - j a_i \bmod d$. One has

$$S_j(x_1, \dots, x_t) = \sum_{\mu \in F_j(x_1,\dots,x_t)} q(x_1 - \mu a_1, x_2 - \mu a_2, \dots, x_t - \mu a_t).$$

On the other side, by Eq. (4), one easily checks:

$$F_j(x_1, \dots, x_t) = \left\{ j + d\mu \in \mathbb{N} \;\middle|\; 0 \le \mu \le \left\lfloor \frac{\lceil \lambda(x_1, \dots, x_t) \rceil - 1 - j}{d} \right\rfloor \right\}.$$

It is important to remark that, in the formula above, if $\lambda(x_1, \dots, x_t) = 0$, then

$$\frac{\lceil \lambda(x_1, \dots, x_t) \rceil - 1 - j}{d} < 0.$$

In this case, since for any $0 \le j \le d-1$, $|-1-j| \le d$, one has

$$\left\lfloor \frac{-1-j}{d} \right\rfloor = -1.$$

Hence $\lambda(x_1, \dots, x_t) = 0$ implies that $F_j(x_1, \dots, x_t)$ is the empty set and thus $S_j$ is the null map. Let us consider the map

$$S : \mathbb{N}^t \times \{\{-1\} \cup \mathbb{N}\} \longrightarrow \mathbb{N},$$

where $S(x_1, \dots, x_t, x)$ is defined as:

$$\begin{cases} \sum_{\mu=0}^{x} q(x_1 - (j+\mu d)a_1, x_2 - (j+\mu d)a_2, \dots, x_t - (j+\mu d)a_t) & x \ge 0 \\ 0 & x = -1. \end{cases}$$

By the definition of the map $S$, for every $(x_1, \dots, x_t) \in C$, with $x_i \equiv d_i \bmod d$, one has:

$$S_j(x_1, \dots, x_t) = S\left(x_1, \dots, x_t, \left\lfloor \frac{\lceil \lambda(x_1, \dots, x_t) \rceil - 1 - j}{d} \right\rfloor\right). \tag{6}$$

Therefore, by applying Lemma 1 to $S$, there exists a polynomial $Q(x_1, \dots, x_t, x)$ such that, for any $(x_1, \dots, x_t, x) \in \mathbb{N}^t \times \{\{-1\} \cup \mathbb{N}\}$,

$$S(x_1, \dots, x_t, x) = Q(x_1, \dots, x_t, x), \tag{7}$$

hence, from Eqs. (6) and (7), for every $(x_1, \ldots, x_t) \in C$, with $x_i \equiv d_i \bmod d$, one has

$$S_j(x_1, \ldots, x_t) = Q(x_1, \ldots, x_t, \left\lfloor \frac{\lceil \lambda(x_1, \ldots, x_t) \rceil - 1 - j}{d} \right\rfloor). \tag{8}$$

By Lemma 16 of the Appendix, one has that:

$$\left\lfloor \frac{\lceil \lambda(x_1, \ldots, x_t) \rceil - 1 - j}{d} \right\rfloor \tag{9}$$

is a quasi-polynomial in the variables $x_1, \ldots, x_t$. Therefore, by Eq. (8) , $S_j$ coincides with a quasi-polynomial on the set of points $(x_1, \ldots, x_t) \in C$, with $x_i \equiv d_i \bmod d$. Obviously this fact holds for any $(d_1, \ldots, d_t) \in \{0, 1, \ldots d-1\}^t$ and, by Lemma 3, $S_j$ is a quasi-polynomial over $C$. Finally, the fact that $F$ is a quasi-polynomial over $C$ follows from Eq. (5) by using Lemma 2. □

**Lemma 5** *Let $\lambda : \mathbb{N}^t \longrightarrow \mathbb{Q}$ be a map such that, for any $(x_1, \ldots, x_t) \in \mathbb{N}^t$,*

$$\lambda(x_1, \ldots, x_t) = b_1 x_1 + \cdots + b_t x_t,$$

*where $b_1, \ldots, b_t$ are given rational coefficients.*

*Let $C, C'$ be subsets of $\mathbb{N}^t$ and let $a_1, \ldots, a_t$ be non negative integers such that the following properties are satisfied: for any $(x_1, \ldots, x_t) \in C$, one has*

- *$\lambda(x_1, \ldots, x_t) \geq 0$,*
- *for any $\lambda \in \mathbb{N}$ such that $\lambda < \lambda(x_1, \ldots, x_t)$, $(x_1 - \lambda a_1, x_2 - \lambda a_2, \ldots, x_t - \lambda a_t) \in C'$.*

*Let $p$ be a quasi-polynomial over $C'$ and define the map $F$ as:*

$$F(x_1, \ldots, x_t) = \sum_{0 \,\leq\, \lambda \,\leq\, \lambda(x_1, \ldots, x_t)} p(x_1 - \lambda a_1, x_2 - \lambda a_2, \ldots, x_t - \lambda a_t).$$

*Then $F$ is a quasi-polynomial over $C$.*

*Proof.* The proof of Lemma 5 is the same of that of Lemma 4 except the point we describe now. In the sum above that defines the map $F$, the index $\lambda$ runs over the set of integers of the closed interval $[0, \lambda(x_1, \ldots, x_t)]$ so that:

$$\lambda \leq \lfloor \lambda(x_1, \ldots, x_t) \rfloor.$$

Therefore, in order to prove the claim, one has to prove a slightly modified version of Eq. (9) of Lemma 4, that is: for any $j = 0, \ldots, d-1$,

$$\left\lfloor \frac{\lfloor \lambda(x_1, \ldots, x_t) \rfloor - j}{d} \right\rfloor,$$

is a quasi-polynomial with rational coefficients in the variables $x_1, \ldots, x_t$. This can be done by using an argument very similar to that one adopted in the proof of Eq. (9). □

**Lemma 6** *Let $\lambda_1, \lambda_2 : \mathbb{N}^t \longrightarrow \mathbb{Q}$ be 2 maps such that, for any $(x_1, \dots, x_t) \in \mathbb{N}^t$,*

$$\lambda_1(x_1, \dots, x_t) = b_1 x_1 + \cdots + b_t x_t, \quad \lambda_2(x_1, \dots, x_t) = c_1 x_1 + \cdots + c_t x_t$$

*where $b_1, \dots, b_t$ and $c_1, \dots, c_t$ are given rational coefficients.*

*Let $C$ be a subset of $\mathbb{N}^t$ and let $a_1, \dots, a_t$ be non negative integers. Suppose that, for any $(x_1, \dots, x_t) \in C$, one has:*

$$0 \le \lambda_1(x_1, \dots, x_t) \le \lambda_2(x_1, \dots, x_t),$$

*and, for any $\lambda \in \mathbb{N}$ such that $\lambda_1(x_1, \dots, x_t) \le \lambda \le \lambda_2(x_1, \dots, x_t)$, one has:*

$$(x_1 - \lambda a_1, x_2 - \lambda a_2, \dots, x_t - \lambda a_t) \in C',$$

*where $C'$ is a given subset of $\mathbb{N}^t$. Let $p$ be a quasi-polynomial over $C'$ and define the map $F$ as:*

$$F(x_1, \dots, x_t) = \sum_{\lambda_1 \le \lambda \le \lambda_2} p(x_1 - \lambda a_1, x_2 - \lambda a_2, \dots, x_t - \lambda a_t),$$

*where $\lambda_1 = \lambda_1(x_1, \dots, x_t)$ and $\lambda_2 = \lambda_2(x_1, \dots, x_t)$. Then $F$ is a quasi-polynomial over $C$. The same result holds whenever the index $\lambda$ runs in the set of integers of the intervals:*

$$(\lambda_1, \lambda_2), \quad (\lambda_1, \lambda_2], \quad [\lambda_1, \lambda_2).$$

*Proof.* Let us solve the case when $\lambda$ runs in the interval $[\lambda_1, \lambda_2]$. Write

$$F(x_1, \dots, x_t) = S_1(x_1, \dots, x_t) - S_2(x_1, \dots, x_t), \tag{10}$$

where:

$$S_1(x_1, \dots, x_t) = \sum_{0 \le \lambda \le \lambda_2} p(x_1 - \lambda a_1, x_2 - \lambda a_2, \dots, x_t - \lambda a_t),$$

and

$$S_2(x_1, \dots, x_t) = \sum_{0 \le \lambda < \lambda_1} p(x_1 - \lambda a_1, x_2 - \lambda a_2, \dots, x_t - \lambda a_t).$$

By applying Lemma 5 to $S_1$ and Lemma 4 to $S_2$, we have that $S_1$ and $S_2$ are quasi-polynomial and by Lemma 2, so is $S_1 - S_2$. The claim now follows from Eq. (10). The other three cases are similarly proved. □

**Lemma 7** *Assuming the same hypotheses of Lemma 4, the function*

$$S(x_1, \dots, x_t) = \sum_{\lambda(x_1, \dots, x_t) \le \lambda \le \lambda(x_1, \dots, x_t)} p(x_1 - \lambda a_1, x_2 - \lambda a_2, \dots, x_t - \lambda a_t).$$

*is a quasi-polynomial over $C$.*

*Proof.* It is a direct consequence of Lemma 6, assuming $\lambda_1(x_1,\ldots,x_t) = \lambda_2(x_1,\ldots,x_t) = \lambda(x_1,\ldots,x_t)$ □

Now we want to define some suitable regions of $\mathbb{R}^t$. More precisely, our regions will be polyhedral cones determined by a family of hyperplanes passing through the origin. We proceed as follows. Let $\pi$ be a plane of $\mathbb{R}^t$. Let us fix an equation for $\pi$ denoted by $\pi(x) = 0$. We associate with $\pi$ a map

$$f_\pi : \mathbb{R}^t \longrightarrow \{+,-,\epsilon\}$$

defined as: for any $x \in \mathbb{R}^t$,

$$f_\pi(x) = \begin{cases} + & \text{if } \pi(x) > 0, \\ \epsilon & \text{if } \pi(x) = 0, \\ - & \text{if } \pi(x) < 0. \end{cases}$$

We remark that the map defined above depends upon the plane $\pi$ and its equation in the obvious geometrical way. We can now give the following important two definitions.

**Definition 3** *Let $\Pi = \{\pi_1,\ldots,\pi_m\}$ be a family of planes of $\mathbb{R}^t$ that satisfy the following property:*

- $\Pi$ *includes the coordinate planes, that is, the planes defined by the equations $x_\ell = 0,\ \ell = 0,\ldots,t$;*
- *every plane of $\Pi$ passes through the origin.*

*Let $\sim$ be the equivalence defined over the set $\mathbb{N}^t$ as: for any $x,\ x' \in \mathbb{N}^t$,*

$$x \ \sim\ x' \quad \Longleftrightarrow \quad \forall\, i = 1,\ldots,m, \quad f_{\pi_i}(x) = f_{\pi_i}(x').$$

*A subset $C$ of $\mathbb{N}^t$ is called a* region (with respect to $\Pi$) *if it is a coset of $\sim$.*

It may be useful to keep in mind that the singleton composed by the origin is a region. Moreover if $t = 2$, the set of all points of $\mathbb{N}^t \setminus \{0\}$ of every line of $\Pi$ is a region also.

**Definition 4** *Let $F : \mathbb{N}^t \longrightarrow \mathbb{N}$ be a map. Then $F$ is said to be a* box spline *in $\mathbb{N}^t$ if there exists a partition $\mathcal{C} = \{C_1,\ldots,C_y\}$ of regions of $\mathbb{N}^t$ – defined by a family of planes satisfying Definition 3 – and a family $p_1,\ldots,p_y$ of quasi-polynomials, every one of which is associated with exactly a region of $\mathcal{C}$, such that, for any $(x_1,\ldots,x_t) \in \mathbb{N}^t$, one has:*

$$F(x_1,\ldots,x_t) = p_a(x_1,\ldots,x_t),$$

*where $a$ is the index of the region $C_a$ that contains $(x_1,\ldots,x_t)$.*

**Lemma 8** *The sum of a finite family of box splines is a box spline.*

*Proof.* It suffices to prove the claim for two box splines. Let $F_1$ and $F_2$ be two box splines and let $\mathcal{C} = \{C_1, \ldots, C_y\}$ and $\mathcal{D} = \{D_1, \ldots, D_z\}$ be the families of regions of $F_1$ and $F_2$ respectively. Moreover, let $\{p_1, \ldots, p_y\}$ and $\{q_1, \ldots, q_z\}$ be the families of quasi polynomials of $F_1$ and $F_2$ respectively.

Consider the box spline defined as follows. Let $\mathcal{E}$ be the partition of regions of $\mathbb{N}^t$ given by the intersection of $\mathcal{C}$ and $\mathcal{D}$ respectively. It is worth noticing that $\mathcal{E}$ is determined by the union of the two families of hyperplanes that define $\mathcal{C}$ and $\mathcal{D}$ respectively. Then we associate the map $r_{lm} = p_l + q_m$ with every region $E_{lm}$ of $\mathcal{E}$. By Lemma 2, $r_{lm}$ is a quasi-polynomial. For any $x \in \mathbb{N}^t$ we have

$$F_1(x) = p_l(x), \quad F_2(x) = q_m(x),$$

where $l$ and $m$ are the indices of the regions $C_l$ and $D_m$ that contain $x$. Hence we have

$$F_1(x) + F_2(x) = r_{lm}(x),$$

while $x$ belongs to the region $E_{lm}$. Since $r_{lm}$ is the quasi polynomial associated with $E_{lm}$, this proves that $F_1 + F_2$ is equal to the box spline defined above. $\square$

The following lemma is a crucial tool in the proof of the main result of this section.

**Lemma 9** *Let $G : \mathbb{N}^t \longrightarrow \mathbb{N}$ be a box spline and let $a_1, \ldots, a_t \in \mathbb{N}$ with $(a_1, ..., a_t) \neq (0, ..., 0)$. Consider the map $\Lambda : \mathbb{N}^t \longrightarrow \mathbb{N}$ that associates with every $(x_1, \ldots, x_t) \in \mathbb{N}^t$, the value*

$$\Lambda(x_1, \ldots, x_t) = \min \left\{ \frac{x_i}{a_i} \mid a_i \neq 0 \right\}.$$

*Let $S : \mathbb{N}^t \longrightarrow \mathbb{N}$ be the map defined as: for every $(x_1, \ldots, x_t) \in \mathbb{N}^t$,*

$$\mathcal{S}(x_1, \ldots, x_t) = \sum_{0 \leq \lambda \leq \Lambda(x_1, \ldots, x_t)} G(x_1 - \lambda a_1, x_2 - \lambda a_2, \ldots, x_t - \lambda a_t). \tag{11}$$

*Then $S$ is a box spline.*

*Proof.* In order to prove the claim, we first associate with the map $S$ a new family of regions that we define now. Let $\Pi$ be the family of planes associated with the box spline $G$. For any $(x_1, x_2, \ldots, x_t) \in \mathbb{N}^t$ consider the line defined by the equation parameterized by $\lambda$:

$$(x_1 - \lambda a_1, x_2 - \lambda a_2, \ldots, x_t - \lambda a_t). \tag{12}$$

Let $\pi$ be a plane of the family $\Pi$ and let $\pi(x) = \sum_{i=1,\ldots,t} \beta_i x_i = 0$ be its equation. The value of $\lambda$ that defines the point of meeting of the line (12) with $\pi$ is easily computed. Indeed, $\lambda$ is such that

$$\sum_{i=1,\ldots,t} \beta_i(x_i - \lambda a_i) = 0,$$

so that

$$\sum_{i=1,\ldots,t} \beta_i x_i = \lambda \cdot \sum_{i=1,\ldots,t} \beta_i a_i \tag{13}$$

which gives

$$\lambda = \sum_{i=1,\ldots,t} \frac{\beta_i}{\gamma} x_i, \tag{14}$$

where

$$\gamma = \sum_{i=1,\ldots,t} \beta_i a_i.$$

It is worth to remark that Eq. (14) is not defined whenever

$$\gamma = \sum_{i=1,\ldots,t} \beta_i a_i = 0. \tag{15}$$

Let us first treat Eq. (15). Here, either the line of Eq. (12) belongs to $\pi$ or such a line is parallel to $\pi$. Therefore, for every point $x$ of the line of Eq. (12), the value of $f_\pi(x)$ is constant so that $\pi$ is not relevant in determining a change of region when a point is moving on the line of Eq. (12). Because of this remark, we shall consider only planes of $\Pi$ for which Eq. (15) does not hold. Denote $\Pi'$ this set of planes. For any $\pi \in \Pi'$, with equation $\pi(x) = \sum_{i=1,\ldots,t} \beta_i x_i = 0$, consider the homogeneous linear polynomial

$$\lambda_\pi(x_1, \ldots, x_t) = \sum_{i=1,\ldots,t} \frac{\beta_i}{\gamma} x_i,$$

where $\gamma = \sum_{i=1,\ldots,t} \beta_i a_i$. We remark that for any $(x_1, x_2, \ldots, x_t) \in \mathbb{N}^t$, the line parameterized by Eq. (12) meets the plane $\pi$ in the point corresponding to the parameter $\lambda = \lambda_\pi(x_1, \ldots, x_t)$.

Take an arbitrary strict total order $<$ on the set $\Pi$ and consider the new family $\widehat{\Pi}$ of planes defined by the following sets of equations:

1. $\pi(x) = 0$, $\pi \in \Pi$
2. $\lambda_{\pi\pi'}(x_1, \ldots, x_k) = 0$, with $\pi, \pi' \in \Pi'$, $\pi < \pi'$, and $\lambda_{\pi\pi'}(x_1, \ldots, x_k) = \lambda_\pi(x_1, \ldots, x_k) - \lambda_{\pi'}(x_1, \ldots, x_k)$.

Call $\widehat{\mathcal{C}}$ the family of regions of $\mathbb{N}^t$ defined by $\widehat{\Pi}$.

We now associate with every region of $\widehat{\mathcal{C}}$ a quasi-polynomial. In order to do this, we need to establish some preliminary facts. Let us fix now a region $C$ of $\widehat{\mathcal{C}}$ and let $x = (x_1, \ldots, x_t)$ be a point of $\mathbb{N}^t$ that belongs to $C$. Let $i$ be such that

$$\Lambda(x) = \frac{x_i}{a_i}.$$

Observe that, for any other point $x' = (x'_1, \ldots, x'_t)$ in $C$, one has

$$\Lambda(x') = \frac{x'_i}{a_i}.$$

Indeed, it is enough to prove that, for any given pair of distinct indices $i, j$, we have:

$$\frac{x_i}{a_i} \leq \frac{x_j}{a_j} \iff \frac{x'_i}{a_i} \leq \frac{x'_j}{a_j}.$$

This is equivalent to say that:

$$\lambda_{\pi\pi'}(x_1, \ldots, x_t) \leq 0 \iff \lambda_{\pi\pi'}(x'_1, \ldots, x'_t) \leq 0,$$

where $\pi, \pi'$ are the planes $x_i = 0$ and $x_j = 0$ respectively. The previous equivalence is true because $x$ and $x'$ belong to the same region of $\widehat{\mathcal{C}}$.

Another important fact is the following. Let us consider any point $x$ of the region $C$ of $\widehat{\mathcal{C}}$. Consider the subset of planes of $\Pi'$:

$$\{\pi_1, \ldots, \pi_m\} = \{\pi \in \Pi' \mid 0 \leq \lambda_\pi(x) \leq \Lambda(x)\}.$$

We can always assume, possibly changing the enumeration of the above planes, that

$$0 \leq \lambda_{\pi_1}(x) \leq \cdots \leq \lambda_{\pi_m}(x) \leq \Lambda(x).$$

**Remark.** Observe that, for any other point $x'$ of $C$, one has

$$\{\pi_1, \ldots, \pi_m\} = \{\pi \in \Pi' \mid 0 \leq \lambda_\pi(x') \leq \Lambda(x)\}.$$

and

$$0 \leq \lambda_{\pi_1}(x') \leq \cdots \leq \lambda_{\pi_m}(x') \leq \Lambda(x').$$

The remark above can be proved by using an argument very similar to that used to prove the previous condition. We suppose that the above inequalities are strict, i.e. $0 < \lambda_{\pi_1}(x) < \cdots < \lambda_{\pi_m}(x) < \Lambda(x)$. In this case, as before, one proves that the same inequalities are strict for any other point $x'$ of the region $C$. The case when the inequalities are not all strict can be treated similarly.

From now on, by the sake of clarity, for any $x = (x_1, \ldots, x_t) \in \mathbb{N}$ we set $y_\lambda(x) = (x_1 - \lambda a_1, \ldots, x_t - \lambda a_t)$.
Consider the following sets:

- $Y_0(x) = \{y_\lambda(x) \mid \lambda \in \mathbb{N} \cap [0, \lambda_{\pi_1}(x))\}$,
- $Y_m(x) = \{y_\lambda(x) \mid \lambda \in \mathbb{N} \cap (\lambda_{\pi_m}, \Lambda(x))\}$,
- $Y_i(x) = \{y_\lambda(x) \mid \lambda \in \mathbb{N} \cap (\lambda_{\pi_i}(x), \lambda_{\pi_{i+1}}(x))\}$, $i = 1, \ldots, m-1$.
- $Z_i(x) = \{y_\lambda(x) \mid \lambda \in \mathbb{N} \cap \{\lambda_{\pi_i}(x)\}\}$, $i = 1, \ldots, m$.
- $Z_{m+1}(x) = \{y_\lambda(x) \mid \lambda \in \mathbb{N} \cap \{\Lambda(x)\}\}$.

We are now able to associate a quasi-polynomial with the region $C$ of $\widehat{\mathcal{C}}$. For this purpose, take two points $x, x'$ in $C$. By the facts discussed before, one has that the lines of Eq. (12) associated with $x$ and $x'$ respectively, meet the planes of $\Pi'$ in the same order. We recall, that a change of region on the generic line of Eq. (12) happens only when the line meets a plane of $\Pi'$. Therefore, since $\widehat{\mathcal{C}}$ is a refinement of $\mathcal{C}$ and $x$ and $x'$ are in a same region with respect to $\mathcal{C}$, the above conditions imply that, for every $i = 0, \ldots, m$, the two sets of points $Y_i(x)$ and $Y_i(x')$ are subsets of a same common region of $\mathcal{C}$. Hence there exists a quasi-polynomial $p_i$, depending on $i$ and on the region $C$, such that, for any $y \in Y_i(x)$ and for any $y' \in Y_i(x')$, $G(y) = p_i(y), G(y') = p_i(y')$. By the previous remark and by Lemma 6, one has that, for any $i = 0, ..., m$, there exists a quasi-polynomial $q_i$, depending on $i$, and on $C$, such that for any $x \in C$

$$q_i(x) = \sum_{y \in Y_i(x)} G(y).$$

Observe that, since $x$ and $x'$ are in the same region $C$, as before one derives that $Z_i(x)$ and $Z_i(x')$ are in the same region with respect to $\mathcal{C}$. Therefore, as before, by applying Lemma 7 there exists a quasi-polynomial $r_i$, depending on $i$ and on $C$, such that for any $x \in C$

$$r_i(x) = \sum_{y \in Z_i(x)} G(y).$$

On the other hand, by Eq. (11), we have that, for any $(x_1, \ldots, x_t) \in C$, $\mathcal{S}(x_1, \ldots, x_t)$ is equal to:

$$q_0(x) + r_1(x) + q_1(x) + r_2(x) + q_2(x) + \cdots r_m(x) + q_m(x) + r_{m+1}(x). \tag{16}$$

Thus, $\mathcal{S}(x_1, \ldots, x_t)$ on the region $C$ is represented as a sum of quasi-polynomials. This, together with Lemma 2 applied to Eq. (16) imply that the map $S$ is a quasi-polynomial over every region of $\widehat{\mathcal{C}}$. The proof of the claim is thus complete.

□

**Theorem 1** *Let*

$$\mathcal{S} : \mathbb{N}^t \longrightarrow \mathbb{N}$$

*be the map which counts, for any vector* $(n_1, \ldots, n_t) \in \mathbb{N}^t$*, the number of distinct non negative solutions of a given Diophantine system:*

$$\begin{cases} a_{11}x_1 + a_{12}x_2 + \cdots + a_{1k}x_k = & n_1 \\ a_{21}x_1 + a_{22}x_2 + \cdots + a_{2k}x_k = & n_2 \\ \qquad\qquad \cdot & \cdot \\ \qquad\qquad \cdot & \cdot \\ \qquad\qquad \cdot & \cdot \\ a_{t1}x_1 + a_{t2}x_2 + \cdots + a_{tk}x_k = & n_t. \end{cases} \tag{17}$$

*where the numbers* $a_{ij} \in \mathbb{N}$ *and, for every* $i = 1, \ldots, k$*, there exists* $j = 1, \ldots, t$ *such that* $a_{ij} \neq 0$*. The map* $\mathcal{S}$ *is a box spline. Moreover such box spline can be effectively constructed starting from the coefficients of the system.*

*Proof.* For any vector $(n_1, \ldots, n_t) \in \mathbb{N}^t$, let $\mathrm{Sol}(n_1, \ldots, n_t)$ be the set of the non negative solutions of the Diophantine system (17) and denote by $\mathcal{S} : \mathbb{N}^t \longrightarrow \mathbb{N}$, the map defined as: for any vector $(n_1, \ldots, n_t) \in \mathbb{N}^t$,

$$\mathcal{S}(n_1, \ldots, n_t) = \mathrm{Card}(\mathrm{Sol}(n_1, \ldots, n_t)),$$

that is, it associates with every vector $(n_1, \ldots, n_t)$ the number of non negative distinct solutions of the system (17). Let us prove that the map $\mathcal{S}$ is a box spline. For this purpose, we proceed by induction on the number of unknowns of the system (17). We start by proving the basis of the induction. In this case, our system has one unknown, say $x$, and it can be written as:

$$\begin{cases} a_1 x = & n_1 \\ a_2 x = & n_2 \\ \quad . & \\ \quad . & \\ \quad . & \\ a_t x = & n_t \end{cases}$$

The system has solutions (and, in this case, it is unique) if and only if there exists $\lambda \in \mathbb{N}$ such that:

$$\lambda(a_1, \ldots, a_t) = (\lambda a_1, \ldots, \lambda a_t) = (n_1, \ldots, n_t). \tag{18}$$

Let us consider the line $\ell$ (through the origin) defined by the parametric equation (18). The line $\ell$ can be determined as the intersection of suitable planes through the origin. Let us consider the family of regions defined by the set of these planes together with the coordinate planes. One can easily associate with every region a quasipolynomial. For this purpose, we remark that the set of points of the line $\ell$ with integral coordinates, without the origin, is a region. On this region, the counting function of the system takes the value 0 or 1. Therefore this map coincides with the quasi-polynomial given by $p = 0,\ q = 1$ with the periodical rule $d = lcm\{a_1, \ldots, a_t\}$. To any other region, we associate $p$. The basis of the induction is thus proved.

Let us now prove the inductive step. If $(x_1, \ldots, x_k) \in \mathrm{Sol}(n_1, \ldots, n_t)$, the system (17) can be written as:

$$\begin{cases} a_{12}x_2 + \cdots + a_{1k}x_k = & n_1 - a_{11}x_1 \\ a_{22}x_2 + \cdots + a_{2k}x_k = & n_2 - a_{21}x_1 \\ \quad . & \quad . \\ \quad . & \quad . \\ \quad . & \quad . \\ a_{t2}x_2 + \cdots + a_{tk}x_k = & n_t - a_{t1}x_1. \end{cases} \tag{19}$$

This implies that:

$$n_1 - a_{11}x_1 \geq 0,\ \ n_2 - a_{21}x_1 \geq 0,\ \ n_t - a_{t1}x_1 \geq 0,$$

so that, since $x_1$ must be an integer $\geq 0$, one has:

$$0 \leq x_1 \leq \frac{n_1}{a_{11}}, \;\; 0 \leq x_1 \leq \frac{n_2}{a_{21}}, \quad \ldots, \;\; 0 \leq x_1 \leq \frac{n_t}{a_{t1}},$$

and thus:

$$0 \leq x_1 \leq \Lambda(x_1, \ldots, x_t),$$

where the map $\Lambda : \mathbb{N}^t \longrightarrow \mathbb{N}$ is defined as:

$$\Lambda(x_1, \ldots, x_t) = \min \left\{ \frac{x_i}{a_{i1}} \;\mid\; a_{i1} \neq 0 \right\}. \tag{20}$$

We remark that, since the vector $(a_{11}, a_{21}, \ldots, a_{t1}) \neq (0, 0, \ldots, 0)$, the map $\Lambda$ is well defined. Set $K = \lfloor \Lambda(x_1, \ldots, x_t) \rfloor$. We can write $\mathrm{Sol}(n_1, \ldots, n_t)$ as:

$$\mathrm{Sol}(n_1, \ldots, n_t) = (0 \times \mathrm{Sol}_0) \,\cup\, (1 \times \mathrm{Sol}_1) \,\cup\, \ldots \,\cup\, (K \times \mathrm{Sol}_K), \tag{21}$$

where, for every $i = 0, \ldots, K$, $\mathrm{Sol}_i$ denotes the set of non negative solutions of the Diophantine system:

$$\begin{cases} a_{12}x_2 + \cdots + a_{1k}x_k = & n_1 - a_{11}i \\ a_{22}x_2 + \cdots + a_{2k}x_k = & n_2 - a_{21}i \\ \qquad\qquad \cdot & \quad \cdot \\ \qquad\qquad \cdot & \quad \cdot \\ \qquad\qquad \cdot & \quad \cdot \\ a_{t2}x_2 + \cdots + a_{tk}x_k = & n_t - a_{t1}i. \end{cases} \tag{22}$$

By Eq. (21), for any $(n_1, \ldots, n_t) \in \mathbb{N}^t$, we have:

$$\mathcal{S}(n_1, \ldots, n_t) = \sum_{i=0,\ldots,K} \mathrm{Card}(\mathrm{Sol}_i). \tag{23}$$

By applying the inductive hypothesis to the system (22), we have that there exists a box spline $G : \mathbb{N}^t \longrightarrow \mathbb{N}$ such that, for any $(n_1, \ldots, n_t) \in \mathbb{N}^t$, if $0 \leq i \leq K$,

$$\mathrm{Card}(\mathrm{Sol}_i) = G(n_1 - a_{11}i, n_2 - a_{21}i, \ldots, n_t - a_{t1}i), \tag{24}$$

so that, by Eq. (23) and Eq. (24), one has:

$$\mathcal{S}(n_1, \ldots, n_t) = \sum_{0 \leq \lambda \leq \Lambda(x_1, \ldots, x_t)} G(n_1 - \lambda a_{11}, n_2 - \lambda a_{21}, \ldots, n_t - \lambda a_{t1}). \tag{25}$$

By Eq. (25), the fact that $\mathcal{S}$ is a box spline follows from Lemma 9. Finally we remark that the proof gives an effective procedure to construct the claimed box spline that describes the map $\mathcal{S}$. □

**Corollary 1** *Let*

$$\mathcal{S} : \mathbb{N}^t \longrightarrow \mathbb{N}$$

*be the map which counts, for any vector $(n_1, \ldots, n_t) \in \mathbb{N}^t$, the number of distinct solutions of a given Diophantine system:*

$$\begin{cases} a_{10} + a_{11}x_1 + a_{12}x_2 + \cdots + a_{1k}x_k = & n_1 \\ a_{20} + a_{21}x_1 + a_{22}x_2 + \cdots + a_{2k}x_k = & n_2 \\ \cdot & \cdot \\ \cdot & \cdot \\ \cdot & \cdot \\ a_{t0} + a_{t1}x_1 + a_{t2}x_2 + \cdots + a_{tk}x_k = & n_t. \end{cases} \tag{26}$$

*where the numbers $a_{ij} \in \mathbb{N}$ and, for every $i = 1, \ldots, k$, there exists $j = 1, \ldots, t$ such that $a_{ij} \neq 0$. Then there exists a finite set $X$ of vectors of $\mathbb{N}^t$ such that the map $\mathcal{S}$ is a box spline over the set $\mathbb{N}^t \setminus X$. Moreover such box spline and the set $X$ can be effectively constructed starting from the coefficients of the system.*

*Proof.* First consider the system

$$\begin{cases} a_{11}x_1 + a_{12}x_2 + \cdots + a_{1k}x_k = & n_1 \\ a_{21}x_1 + a_{22}x_2 + \cdots + a_{2k}x_k = & n_2 \\ \cdot & \cdot \\ \cdot & \cdot \\ \cdot & \cdot \\ a_{t1}x_1 + a_{t2}x_2 + \cdots + a_{tk}x_k = & n_t. \end{cases} \tag{27}$$

According to Theorem 1, there exists a box spline $F$ that counts, for every $(n_1, \ldots, n_t) \in \mathbb{N}^t$, the number of the solutions of the diophantine system (27). Let $\mathcal{C} = \{C_1, \ldots, C_y\}$ be the family of partitions of $\mathbb{N}^t$ and $\{p_1, \ldots, p_m\}$ be the family of quasi-polynomials that define $F$.

Let $a_0 = (a_{10}, \ldots, a_{t0})$ be the vector whose components are the constants $a_{i0}$ of the system (26) and let $X$ be the subset of vectors $\eta$ of $\mathbb{N}^t$ such that $\eta - a_0 \notin \mathbb{N}^t$. The subset $X$ is finite. For every $\eta \in \mathbb{N}^t \setminus X$, one has that:

$$\mathcal{S}(\eta) = p_z(\eta - a_0),$$

where $z$ is the index of the region of the family $\mathcal{C}$ that contains the vector $\eta - a_0$. Therefore, the quasi polynomial used to compute the map $\mathcal{S}$ at the point $n$ is obtained, from one of $F$, by a traslation, modulo the vector $a_0$, of the vector $\eta$.

From the argument above, one can easily show that the map $\mathcal{S}$ is a box spline. □

# 3 Preliminaries on context-free languages

## 3.1 Semi-linear and semi-simple sets

The aim of this paragraph is to recall some classical results about rational sets of the free commutative monoid. We follow the notation adopted in [23]. Let $M$ be an additive commutative monoid and let $B = \{b_1, \ldots, b_k\}$ be a finite subset of $M$. Then we denote by $B^{\oplus}$ the submonoid generated by $B$, that is

$$B^{\oplus} = \{n_1 b_1 + \cdots + n_k b_k \mid n_i \geq 0\}.$$

The following definitions are useful.

**Definition 5** *A subset $X$ of a commutative monoid $M$ is:*

1. linear *if $X = x + B^{\oplus}$ where $x \in M$ and $B$ is finite;*
2. simple *if $X$ is linear and $X = x + B^{\oplus}$ where $B^{\oplus}$ is a free commutative monoid with basis $B$ and the sum $x + B^{\oplus}$ is unambiguous;*
3. semi-linear *if $X$ is a finite union of linear sets;*
4. semi-simple *if $X$ is a finite disjoint union of simple sets.*

**Remark 1** In the definition of simple set, the vector $x$ and those of $B$ shall be called a *representation* of $X$.

In a commutative monoid, semi-linear sets are obviously rational. Conversely, the following result, due to Eilenberg and Schützenberger ([7]), allows to prove a remarkable property of rational sets (*see* [23] for a proof).

**Theorem 2** *The rational sets of a commutative monoid are semi-simple.*

**Remark 2** Theorem 2 is effective for free commutative monoids. Indeed, starting from a rational set $X$, one can effectively represent $X$ as a semi-linear set. Moreover, starting from a semi-linear set $X$, one can effectively construct a finite family of finite disjoint sets of vectors, each one generating a simple set, and such that their union is $B$.

The aim of this paragraph is to recall some classical results about context-free and regular languages (*see* [1, 6, 14, 23]). Now we recall the celebrated *Cross-Section theorem* by Eilenberg.

**Theorem 3** *Let $\alpha : A^* \longrightarrow B^*$ be a morphism and let $L$ be a rational language of $A^*$. Then one can effectively construct a rational subset $L'$ of $L$ such that $\alpha$ maps bijectively $L'$ onto $\alpha(L)$.*

Let $A = \{a_1, \ldots, a_t\}$ be a finite alphabet and $u \in A^*$ be a word. Then the *Parikh vector* of $u$ is defined as

$$\psi(u) = (|u|_{a_1}, \ldots, |u|_{a_t}),$$

and the map

$$\psi : A^* \longrightarrow \mathbb{N}^t,$$

defined above is the canonical epimorphism associated with the free commutative monoid $\mathbb{N}^k$. In the sequel, $\psi$ will be also called the *Parikh map*. Now we state the following well known theorem due to Parikh.

**Theorem 4** *The image of any context-free language under the Parikh map is an effective semi-linear set.*

## 3.2 Bounded languages

The aim of this paragraph is to present some results concerning bounded context-free languages. Let us first introduce the notion of bounded language.

**Definition 6** *Let $L$ be a language of $A^*$. Then, for any positive integer $n$, $L$ is called $n$-*bounded *if there exist nonempty words $u_1, \ldots, u_n \in A^*$ such that*

$$L \subseteq u_1^* \cdots u_n^*.$$

*Moreover, we say that $L$ is* bounded *if there exists an integer $n$ such that $L$ is $n$-bounded.*

**Theorem 5** *(Ginsburg, [9]) It is decidable whether a context-free language is bounded or not.*

**Remark 3** The procedure involved in the test of Theorem 5 allows to construct, from a given bounded context-free language $L$, a finite set $\{u_1, \ldots, u_k\}$ of words such that $L \subseteq u_1^* \cdots u_k^*$.

Let us consider a bounded language $L \subseteq u_1^* \cdots u_k^*$. We set

$$Ind(L) = \{(l_1, \ldots, l_k) \in \mathbb{N}^k \ \mid \ u_1^{l_1} \cdots u_k^{l_k} \in L\}.$$

The following result was proven in [9]. For the sake of completeness, we give a simple constructive proof using Theorem 4.

**Theorem 6** *Let $L$ be a bounded context-free language. Then $Ind(L)$ is a semi-linear set. Moreover, one can effectively construct $Ind(L)$.*

*Proof.* Let $\Sigma$ be the alphabet of $L$ and let $L \subseteq u_1^* \cdots u_k^*$. Let $A = \{a_1, \ldots, a_k\}$ be a new alphabet with $k$ letters. Consider the morphism

$$\zeta : A^* \longrightarrow \Sigma^*, \tag{28}$$

generated by the map,

$$\forall\, i = 1, \ldots, k, \qquad a_i \ \longrightarrow \ u_i.$$

Since $L$ is context-free, the language

$$X = \zeta^{-1}(L) \ \cap \ a_1^* \cdots a_k^*,$$

is also context-free and by Theorem 4, $\psi(X)$ is semi-linear. Finally it is easily seen that $Ind(L) = \psi(X)$. Indeed, for every vector $x = (l_1, \ldots, l_k) \in \mathbb{N}^k$, we have,

$x \in Ind(L) \implies u_1^{l_1} \cdots u_k^{l_k} \in L \implies a_1^{l_1} \cdots a_k^{l_k} \in X \implies$
$\implies \psi(a_1^{l_1} \cdots a_k^{l_k}) = x \in \psi(X),$

so that $Ind(L) \subseteq \psi(X)$. The inverse inclusion is similarly proved. Hence $Ind(L)$ is semi-linear.

Since every step of this proof and Theorem 4 are effective, one has that $Ind(L)$ can be effectively computed starting from $L$. □

If $L \subseteq u_1^* \cdots u_k^*$, then we define the map:

$$\phi : \mathbb{N}^k \longrightarrow u_1^* \cdots u_k^*, \tag{29}$$

such that, for every vector $(l_1, \ldots, l_k) \in \mathbb{N}^k$,

$$\phi((l_1, \ldots, l_k)) = u_1^{l_1} \cdots u_k^{l_k}.$$

The following result proved in [12] is a consequence of Theorems 3 and 4.

**Lemma 10** *Let $L \subseteq u_1^* \cdots u_k^*$ be a bounded context-free language. Then there exists a semi-linear set $B$ of $\mathbb{N}^k$ such that $\phi(B) = L$ and $\phi$ is injective on $B$. Moreover, $B$ can be effectively constructed.*

*Proof.* Let $\Sigma$ be the alphabet of $L$ and $A = \{a_1, \ldots, a_k\}$ be an alphabet with $k$ letters. Consider now the morphism $\zeta : A^* \longrightarrow \Sigma^*$ as defined in (28). Since

$$\zeta(a_1^* \cdots a_k^*) = u_1^* \cdots u_k^*,$$

by Theorem 3, there exists a regular subset $R$ of $a_1^* \cdots a_k^*$ such that $\zeta$ maps bijectively $R$ onto $u_1^* \cdots u_k^*$. Let $L'$ be the language defined as

$$L' = \zeta^{-1}(L) \cap R. \tag{30}$$

Since $L'$ is context-free, by Theorem 6, the set $Ind(L')$ is a semi-linear set of $\mathbb{N}^k$. Set $B = Ind(L')$. As shown in [12], one can easily prove that $L = \phi(B)$ and, moreover, $\phi$ is injective on $B$.

Let us finally prove that $B$ is constructible. Indeed, by Theorem 3, the set $R$ is effectively constructible. On the other hand, by applying standard results, the set $L'$ defined in Eq. (30) is an effective context-free language. By using Theorem 6, we can effectively construct the set $B = Ind(L')$ which is semi-linear. □

We finally close this paragraph by stating the following remarkable characterization of bounded context-free languages. (*see* [15, 21, 22]).

**Theorem 7** *Let $L$ be a context-free language. Then $L$ is sparse if and only if $L$ is bounded.*

# 4 On the Parikh map of bounded context-free languages

The first result we prove, concerns the structure of the Parikh map of a bounded context-free language. More precisely, we will prove that, given a sparse context-free language $L$ on an alphabet $A = \{a_1, \ldots, a_t\}$, it is possible to effectively associate with $L$ a box spline $F : \mathbb{N}^t \longrightarrow \mathbb{N}$ such that, for any vector $(n_1, \ldots, n_t) \in \mathbb{N}^t$, one has:

$$\psi(n_1, \ldots, n_t) = F(n_1, \ldots, n_t).$$

In the sequel, we make the following assumption.

**Assumption** We assume that $L \subseteq u_1^* \cdots u_k^*$ is a bounded context-free language and, according to Lemma 10 and Theorem 2, there exists a semi-simple set $B$ such that $L = \phi(B)$ and $\phi$ is injective on $B$. Set

$$B = \bigcup_{i=1,\ldots,s} B_i, \tag{31}$$

where, for every $i = 1, \ldots, s$, $B_i$ is simple and let

$$L = \bigcup_{i=1,\ldots,s} L_i, \tag{32}$$

where, for every $i = 1, \ldots, s$, $L_i = \phi(B_i)$.
We need some preliminary results.

**Lemma 11** *Let $L_i$ and $L_j$ be two languages of Eq. (32) with $i \neq j$. Then $L_i$ and $L_j$ are disjoint.*

*Proof.* By contradiction, suppose that $L_i \cap L_j \neq \emptyset$ and let $x \in L_i \cap L_j$.Then there exist $c_i \in B_i$ and $c_j \in B_j$ such that

$$x = \phi(c_i) = \phi(c_j).$$

By the injectivity of $\phi$ on $B$, we have

$$c_i = c_j$$

and thus

$$B_i \cap B_j \neq \emptyset,$$

which is a contradiction. This proves the claim. □

**Lemma 12** *Let $B_i$ be a simple set of Eq. (31) and*

$$B_i = b_0 + b_1^{\oplus} + \cdots + b_n^{\oplus},$$

*where $b_0, \ldots, b_n$ are the vectors of the representation of $B_i$. Then, for every vector $v = (v_1, \ldots, v_t) \in \mathbb{N}^t$, the number of words of $L_i$ whose Parikh vector is*

*$v$, equals the number of non-negative integer solutions of the Diophantine system of the $t$ equations $\{E_\ell(v)\}_{\ell=1,\ldots,t}$ :*

$$\left\{\begin{array}{rl} \lambda_0^1+\lambda_1^1 x_1+\lambda_2^1 x_2+\cdots+\lambda_n^1 x_n = & v_1 \\ \lambda_0^2+\lambda_1^2 x_1+\lambda_2^2 x_2+\cdots+\lambda_n^2 x_n = & v_2 \\ \cdot & \cdot \\ \cdot & \cdot \\ \cdot & \cdot \\ \lambda_0^t+\lambda_1^t x_1+\lambda_2^t x_2+\cdots+\lambda_n^t x_n = & v_t \\ \cdot & \end{array}\right. \tag{33}$$

*where, for every $i=0,\ldots,n$ and $j=1,\ldots,t$,*

$$\lambda_i^j = |\phi(b_i)|_{a_j}.$$

*Proof.* For any vector $v=(v_1,\ldots,v_t)\in\mathbb{N}^t$, let $S_v$ be the subset of vectors $x=(x_1,\ldots,x_n)\in\mathbb{N}^n$ which are solutions of the system (33). Define the map:

$$\theta : S_v \longrightarrow L_i,$$

as

$$\theta(x)=\theta(x_1,\ldots,x_n)=\phi(b_0+b_1x_1+\cdots+b_nx_n).$$

One can check that, for every $\ell=1,\ldots,t$:

$$|\phi(b_0+b_1x_1+\cdots+b_nx_n)|_{a_\ell}=\lambda_0^\ell+\lambda_1^\ell x_1+\cdots+\lambda_n^\ell x_n=v_\ell,$$

so that the codomain of $\theta$ is $L_i\cap\psi^{-1}(v)$, that is the set of all words of $L_i$ whose Parikh vector is $v$.
Now we prove that $\theta$ is a bijection of $S_v$ onto the language $L_i\cap\psi^{-1}(v)$. The map $\theta$ is injective on its domain. Indeed, let $x=(x_1,\ldots,x_n),\ y=(y_1,\ldots,y_n)\in S_v$. If $\theta(x)=\theta(y)$ then

$$\phi(b_0+b_1x_1+\cdots+b_nx_n)=\phi(b_0+b_1y_1+\cdots+b_ny_n),$$

and, by the injectivity of $\phi$ on $B_i$, we have

$$b_0+b_1x_1+\cdots+b_nx_n=b_0+b_1y_1+\cdots+b_ny_n.$$

Since $B_i$ is simple, the latter gives

$$\forall\ i=1,\ldots,n,\ \ x_i=y_i,$$

thus obtaining $x=y$.

We prove that the map $\theta$ is surjective. Indeed, let $u\in L_i\cap\psi^{-1}(v)$ and let $x\in B_i$ such that $\phi(x)=u$. Write $x$ as $x=b_0+b_1x_1+\cdots+b_nx_n$. One has $v=\psi(u)$. It is easily checked that, for every $\ell=1,\ldots,t$:

$$\lambda_0^\ell+\lambda_1^\ell x_1+\cdots+\lambda_n^\ell x_n=v_\ell.$$

Hence so that $x\in S_v$ and $\theta(x)=u$. Thus $\theta$ is surjective and, therefore, $\mathrm{Card}(L_i\cap\psi^{-1}(v))=\mathrm{Card}(S_v)$. The proof of the lemma is thus complete. □

**Lemma 13** *Let $L_i$ be a language of Eq. (32). Then there exists a box spline $F : \mathbb{N}^t \longrightarrow \mathbb{N}$ and a finite set $X_{L_i}$ of vectors of $\mathbb{N}^t$ such that, for any vector*

$$(n_1, \ldots, n_t) \in \mathbb{N}^t \ \setminus \ X_{L_i},$$

*one has:*

$$F(n_1, \ldots, n_t) = \mathrm{Card}(\{u \in L_i \ | \ \psi(u) = (n_1, \ldots, n_t)\}).$$

*Proof.* It immediately follows from Lemma 12 and Corollary 1. □

**Theorem 8** *Let $L$ be the language of Eq. (32). Then there exist a box spline $F : \mathbb{N}^t \longrightarrow \mathbb{N}$ and a finite set $X_L$ of vectors of $\mathbb{N}^t$ such that, for any vector*

$$(n_1, \ldots, n_t) \in \mathbb{N}^t \ \setminus \ X_L,$$

*one has:*

$$F(n_1, \ldots, n_t) = \mathrm{Card}(\{u \in L \ | \ \psi(u) = (n_1, \ldots, n_t)\}).$$

*Proof.* For the sake of simplicity, assume that $L = L_1 \cup L_2$, the proof in the general case being completely similar. By Lemma 11, we have that, for every $\eta = (n_1, \ldots, n_t) \in \mathbb{N}^t$, the number

$$c(\eta) = \mathrm{Card}(\{u \in L \ | \ \psi(u) = n\})$$

is equal to:

$$c(\eta) = c_1(\eta) + c_2(\eta), \tag{34}$$

where

$$c_1(\eta) = \mathrm{Card}(\{u \in L_1 \ | \ \psi(u) = \eta\}), \quad c_2(\eta) = \mathrm{Card}(\{u \in L_2 \ | \ \psi(u) = \eta\}).$$

By Lemma 13, there exist two finite subsets $X_1, \ X_2$ of $\mathbb{N}^t$ and two box splines $F_1, F_2$ such that:

$$\forall \, \eta \in \mathbb{N}^t \ \setminus \ X_{L_1}, \ \ c_1(\eta) = F_1(\eta),$$

and

$$\forall \, \eta \in \mathbb{N}^t \ \setminus \ X_{L_2}, \ \ c_2(\eta) = F_2(\eta).$$

Let $X_L = X_1 \cup X_2$ and let $F = F_1 + F_2$. For any $\eta \in \mathbb{N}^t \ \setminus \ X_L$, we therefore have

$$c(\eta) = c_1(\eta) + c_2(\eta) \ = \ F_1(\eta) + F_2(\eta) \ = \ F(\eta).$$

The claim finally follows by observing that, by Lemma 8, $F$ is a box spline. □

**Theorem 9** *The box spline $F$ and the finite set $X_L$ defined in the statement of Theorem 8 can be effectively constructed starting from the language $L$.*

*Proof.* The proof is a walk through the results, each one being effective, we gathered so far. It is useful to divide the proof into the following subsequent steps.

**Step 1.** Starting from $L$, one can effectively construct a finite set $\{u_1, \ldots, u_k\}$ of nonempty words such that $L \subseteq u_1^* \cdots u_k^*$. This is done by executing the procedure involved in Theorem 5 (*cf.* Remark 3).

**Step 2.** One can effectively construct a semi-linear set $B \subseteq \mathbb{N}^k$ such that $L = \phi(B)$ and $\phi$ is injective on $B$. This is done in Lemma 10.

**Step 3.** One can effectively represent $B$ as a semi-simple set. More precisely, one can construct a finite family of finite sets of vectors, say $\{V_i\}$, where $V_i$ is a representation of $B_i$ and such that the union of the $B_i$'s is $B$. This is done according to Theorem 2 and Remark 2.

**Step 4.** For every $n \geq 0$ and for every set $V_i$, one can effectively construct the Diophantine system of the $t$ equations $\{E_\ell(v)\}_{\ell=1,\ldots,t}$ stated in Lemma 12. This is done by using the sets of words $\{u_1, \ldots, u_k\}$ and the vectors of $V_i$.

**Step 5.** Since Corollary 1 is constructive, for every language $L_i = \phi(B_i)$, one can effectively construct a box spline $F_i : \mathbb{N}^t \longrightarrow \mathbb{N}$ and a finite set $X_{L_i}$ of vectors of $\mathbb{N}^t$ such that, for any vector

$$(n_1, \ldots, n_t) \in \mathbb{N}^t \setminus X_{L_i},$$

one has:

$$F_i(n_1, \ldots, n_t) = \mathrm{Card}(\{u \in L_i \mid \psi(u) = (n_1, \ldots, n_t)\}).$$

Finally, by applying the construction shown in the proof of Theorem 8 and starting from the set of box splines defined in Step 5 for every language $L_i$, we can effectively obtain the box spline $F$ considered in the claim. □

# 5 On the rationality of the Parikh map of a sparse context-free language

As shown in Section 4, given a bounded context-free language, one can effectively construct a Diophantine system such that its counting function associates with every vector of $\mathbb{N}^t$ the number of words of the language whose images, under the Parikh map, is the given vector. The aim of this section is to prove that such a map is rational. In order to prove this result, some preliminary notions and results concerning formal power series have to be recalled. We follow the classical reference [20]. Let $\mathbb{K}$ be a commutative semiring and let $X = \{x_1, \ldots, x_t\}$ be a set of $t$ commutative variables. We identify the set of all commutative monomials over $X$ with $\mathbb{N}^t$. We denote by $\mathbb{K}[X]$ and by $\mathbb{K}[[X]]$ respectively the semiring of polynomials and the semiring of formal power series on the set of commutative variables $X$ and with coefficients taken in $\mathbb{K}$. A power series is a map $\mathbb{N}^t \longrightarrow \mathbb{K}$. Any power series $r$ of $\mathbb{K}[[X]]$ can be written as a formal sum

$$r = \sum_{n_1,\ldots,n_t \in \mathbb{N}} (r, x_1^{n_1} \cdots x_t^{n_t}) x_1^{n_1} \cdots x_t^{n_t},$$

where $(r, x_1^{n_1} \cdots x_t^{n_t})$ is the coefficient of $\mathbb{K}$ associated with the monomial $x_1^{n_1} \cdots x_t^{n_t}$ by the series $r$. We recall that the family of rational power series of $\mathbb{K}[[X]]$, denoted $Rat(\mathbb{K}[[X]])$, is the smallest subset of $\mathbb{K}[[X]]$ that contains $\mathbb{K}[X]$ and that is closed with respect to the rational operations, that is, the operations that, given series $s, t \in \mathbb{K}[[X]]$, associate with them, the sum $s + t$, the (Cauchy) product $st$ and the star $s^* = \sum_{i=0}^{\infty} s^i$. We will prove the following statement.

**Theorem 10** *Let us consider a Diophantine system defined as:*

$$\left\{ \begin{array}{rl} a_{01} + a_{11}x_1 + a_{12}x_2 + \cdots + a_{1k}x_k = & n_1 \\ a_{02} + a_{12}x_1 + a_{22}x_2 + \cdots + a_{2k}x_k = & n_2 \\ \cdot & \cdot \\ \cdot & \cdot \\ \cdot & \cdot \\ a_{0t} + a_{1t}x_1 + a_{2t}x_2 + \cdots + a_{tk}x_k = & n_t \\ \cdot & \end{array} \right. \tag{35}$$

*where, for every $i = 0, \ldots, k$ and $j = 1, \ldots, t$, $a_{ji} \in \mathbb{N}$. Let $\mathcal{S} : \mathbb{N}^t \longrightarrow \mathbb{N}$ be the counting function of the system, that is the map that associates with every $(n_1, \ldots, n_t) \in \mathbb{N}^t$ the number of non negative solutions of the system. Then $\mathcal{S}$ is $\mathbb{N}$-rational.*

Let us associate with the system (35) a formal power series $S \in \mathbb{N}[[X]]$ defined as:

$$S = S_0 S_1 \cdots S_k,$$

with

$$S_0 = x_1^{a_{01}} \cdots x_t^{a_{0t}},$$

and, for every $i = 1, \ldots, k$,

$$S_i = (x_1^{a_{i1}} \cdots x_t^{a_{it}})^*,$$

where the numbers $a_{ij}$ are the coefficients of the system (35).

Since, for every $i = 0, \ldots, k$, the series $S_i \in Rat(\mathbb{N}[[X]])$ and since $Rat(\mathbb{N}[[X]])$ is closed under the rational operations, one has that $S \in Rat(\mathbb{N}[[X]])$.

**Lemma 14** *The series $S$ is equal to the map $\mathcal{S}$.*

*Proof.* By developping the formal series $S = S_0 S_1 \cdots S_k$, we have that $S$ is equal to:

$$x_1^{a_{01}} \cdots x_t^{a_{0t}} \cdot \left( \sum_{\mu_1 \geq 0} (x_1^{a_{11}} \cdots x_t^{a_{1t}})^{\mu_1} \right) \cdots \left( \sum_{\mu_k \geq 0} (x_1^{a_{k1}} \cdots x_t^{a_{kt}})^{\mu_k} \right).$$

By the formula above, one can check that, for any $x_1^{n_1} \cdots x_t^{n_t}$, the coefficient $(S, x_1^{n_1} \cdots x_t^{n_t})$ of $x_1^{n_1} \cdots x_t^{n_t}$ is the cardinal number of the set:

$$\{(\mu_1, \ldots, \mu_k) \in \mathbb{N}^t \mid x_1^{n_1} \cdots x_t^{n_t} = x_1^{a_{01}} \cdots x_t^{a_{0t}} \cdot (x_1^{a_{11}} \cdots x_t^{a_{1t}})^{\mu_1} \cdots (x_1^{a_{k1}} \cdots x_t^{a_{kt}})^{\mu_k}\}).$$

The number above is cleary equal to the value of the map $\mathcal{S}$ computed at $(n_1, \ldots, n_t)$ and this concludes the proof. □

Now we can prove Theorem 10.

**Proof of Theorem 10** By Lemma 14, the counting function $\mathcal{S}$ of the Diophantine system (35) coincides with the formal power series $S$. The claim follows by recalling that $S$ is $\mathbb{N}$-rational. □

We can now prove tha announced result of the section. Let $L$ be a given language over the alphabet $A = \{a_1, \ldots, a_t\}$ and let $\phi_L : \mathbb{N}^t \longrightarrow \mathbb{N}$ be the map defined as: for every $x \in \mathbb{N}^t$,

$$\phi_L(x) = \text{Card}(\{u \in L \mid \psi(u) = x\}).$$

**Corollary 2** *The map $\phi_L$ of a bounded context-free language $L$ is $\mathbb{N}$-rational.*

*Proof.* As shown in Section 4, given a bounded context-free language, one can effectively construct a Diophantine system defined as in Eq. (35) such that its counting function coincides with the map $\phi_L$. Then the claim follows by applying Theorem 10. □

**Acknowledgements**. The authors would like to thank Corrado De Concini for very useful comments and discussions concerning the results presented in the first part of the paper.

# References

[1] J. Berstel, *Transductions and Context-Free Languages* Teubner, Stuttgart, 1979.

[2] F. D'Alessandro, B. Intrigila, S. Varricchio, On the structure of the counting function of context-free languages, *Theoret. Comput. Sci.* **356**, 104–117 (2006).

[3] W. Dahmen and C.A. Micchelli, The number of solutions to linear Diophantine equations and multivariate splines, *Trans. Amer. Math. Soc.* **308**, no. 2, 509–532 (1988).

[4] C. De Concini and C. Procesi, The algebra of the box-spline, Preprint arXiv:math.NA/0602019 (2006).

[5] C. De Concini, C. Procesi, and M. Vergne, Partition function and generalized Dahmen-Micchelli spaces, Preprint arXiv:0805.2907 (2008)

[6] A. de Luca and S. Varricchio, *Finiteness and Regularity in Semigroups and Formal Languages*, Springer-Verlag, 1999.

[7] S. Eilenberg and M.-P. Schützenberger, Rational sets in commutative monoids, *Journal of Algebra,* **13 (2)**, 173–191 (1969).

[8] P. Flajolet, Analytic models and ambiguity of context-free languages, *Theoretical Computer Science* **49**, 283–309 (1987).

[9] S. Ginsburg, *The mathematical theory of context-free languages*, Mc Graw-Hill, New York, 1966.

[10] Ronald L. Graham, Donald E. Knuth, and Oren Patashnik, *Concrete mathematics: a foundation for computer science*, Reading, Mass.: Addison-Wesley, 1989.

[11] J. Honkala, A decision method for Parikh slenderness of context-free languages, *Discrete Appl. Math.* **73**, 1–4 (1997).

[12] J. Honkala, Decision problems concerning thinness and slenderness of formal languages, *Acta Informatica* **35**, 625–636 (1998).

[13] J. Honkala, On Parikh slender context-free languages, *Theoretical Computer Science* **255**, 667–677 (2001).

[14] J. Hopcroft and J. Ullman, *Introduction to Automata Theory, Languages and Computation,* Addison-Wesley Pub. Co., 1979.

[15] O. Ibarra and B. Ravikumar, On sparseness, ambiguity and other decision problems for acceptors and transducers, *Lecture Notes in Computer Science*, Vol. 210, pp. 171–179, Springer-Verlag, Berlin, 1986.

[16] L. Ilie, On generalized slenderness of context-free languages, *Words, semigroups, & transductions,* pp. 189–202, World Sci. Publ., River Edge, NJ, 2001.

[17] L. Ilie,G. Rozenberg, and A. Salomaa, A characterization of poly-slender context-free languages, *Theor. Inform. Appl.* **34** no. 1, 77–86 (2000).

[18] R. Incitti, The growth function of context-free languages, *Theoretical Computer Science* **255**, 601–605 (2001).

[19] D. E. Knuth, *The Art of Computer Programming*, vol. 1, Addison-Wesley, 1968.

[20] W. Kuich and A. Salomaa, *Semirings, Automata, Languages*, Springer-Verlag, 1985.

[21] M. Latteux and G. Thierrin, On bounded context-free languages, *Elektron. Informationsverarb. Kybernet.* **20**, 3–8 (1984).

[22] D. Raz, Length considerations in context-free languages, *Theoretical Computer Science* **183**, 21–32 (1997).

[23] J. Sakarovitch, *Éléments de théorie des automates*, Vuibert, Paris, 2003.

[24] A. Szenes and M. Vergne, Residue formulae for vectors partitions and Euler-Maclaurin sums. Formal power series and algebraic combinatorics (Scottsdale, AZ, 2001). *Adv. in Appl. Math.* **30 (1-2)**, 295–342 (2003).

## Appendix of Section 2

The following lemma can be proved by following [19], Ch. 1.

**Lemma 15** *Let $m \in \mathbb{N}$. There exists a polynomial $p$ in one variable $x$ with rational coefficients such that:*

*1) for any $n \in \mathbb{N}$, $p(n) = \sum_{\lambda=0,\dots,n} \lambda^m$,*

*2) $p$ factorizes as $p(x) = (x+1)p'(x)$ where $p'$ is a polynomial in one variable $x$ with rational coefficients.*

*Proof.* Let $m \geq 1$. There exist numbers $b_0, \dots, b_m$ such that, for every $k \in \mathbb{N}$, the number $k^m$ can be expressed as:

$$k^m = b_0 \binom{k}{0} + \cdots + b_m \binom{k}{m}.$$

Therefore, the previous equation gives:

$$\sum_{k=0,\dots,n} k^m = b_0 \cdot \sum_{k=0,\dots,n} \binom{k}{0} + b_1 \cdot \sum_{k=0,\dots,n} \binom{k}{1} + \cdots + b_m \cdot \sum_{k=0,\dots,n} \binom{k}{m}. \tag{36}$$

On the other hand, one has that:

$$\sum_{k=0,\dots,n} \binom{k}{r} = \binom{n+1}{r+1}. \tag{37}$$

By applying Eq. (37) to every addendum of the sum of Eq. (36), one obtains a polynomial $p$ that satifies the claim of the lemma. □

**Example** Taking $m = 2$,we express as a polynomial the sum of the squares of the first $k$ non negative integers. Let us recall that, for any $k \in \mathbb{N}$:

$$k^2 = 2\binom{k}{2} + \binom{k}{1}.$$

Then, for any $n \in \mathbb{N}$, by applying Eq. (37), one has:

$$\sum_{k=0,\dots,n} k^2 = 2 \sum_{k=0,\dots,n} \binom{k}{2} + \sum_{k=0,\dots,n} \binom{k}{1} = 2\binom{n+1}{3} + \binom{n+1}{2} =$$

$$= 2\frac{(n+1)n(n-1)}{6} + \frac{(n+1)n}{2}$$

which finally gives the claimed polynomial.

**Lemma 16** *Let $d \in \mathbb{N}$ and let $\lambda : \mathbb{N}^t \longrightarrow \mathbb{Q}$ be the map such that, for any $(x_1, \ldots, x_t) \in \mathbb{N}^t$,*

$$\lambda(x_1, \ldots, x_t) = \sum_1^t b_i x_i,$$

*where $b_1, \ldots, b_t$ are non negative rational numbers. Let $k$ be a constant integer, then the map*

$$\phi(x_1, \ldots, x_t) = \left\lfloor \frac{\lceil \lambda(x_1, \ldots, x_t) \rceil + k}{d} \right\rfloor.$$

*is a quasi-polynomial.*

*Proof.* We can represent the rational numbers $b_1, \ldots, b_t$ as $b_1 = f_1/g, \ \ldots, b_t = f_t/g$, where $f_1, \ldots, f_t$ and $g$ are non negative integers.

Let $(x_1, \ldots, x_t) \in \mathbb{N}^t$. For every $i = 1, \ldots, t$, let $r_i, \alpha_i \in \mathbb{N}$ such that:

$$x_i = \alpha_i g d \ + \ r_i, \quad 0 \leq r_i < gd. \tag{38}$$

By using Eq. (38), one derives

$$\phi(x_1, \ldots, x_t) = \left\lfloor \frac{\left\lceil d \sum_{i=1,\ldots,t} f_i \alpha_i + \sum_{i=1,\ldots,t} \frac{f_i r_i}{g} \right\rceil + k}{d} \right\rfloor =$$

$$= \left\lfloor \frac{d \left( \sum_{i=1,\ldots,t} f_i \alpha_i \right) + \left\lceil \sum_{i=1,\ldots,t} \frac{f_i r_i}{g} \right\rceil + k}{d} \right\rfloor = \left\lfloor \left( \sum_{i=1,\ldots,t} f_i \alpha_i \right) + \frac{\left\lceil \sum_{i=1,\ldots,t} \frac{f_i r_i}{g} \right\rceil + k}{d} \right\rfloor =$$

$$= \left( \sum_{i=1,\ldots,t} f_i \alpha_i \right) + \left\lfloor \frac{\left\lceil \sum_{i=1,\ldots,t} \frac{f_i r_i}{g} \right\rceil + k}{d} \right\rfloor = \left( \sum_{i=1,\ldots,t} \frac{b_i}{d} (x_i - r_i) \right) + \left\lfloor \frac{\left\lceil \sum_{i=1,\ldots,t} \frac{f_i r_i}{g} \right\rceil + k}{d} \right\rfloor.$$

For any $r_1, \ldots, r_t$, with $0 \leq r_i \leq gd - 1$, consider the polynomial

$$p_{(r_1,\ldots,r_t)}(x_1, \ldots, x_t) = \left( \sum_{i=1,\ldots,t} \frac{b_i}{d} (x_i - r_i) \right) + \left\lfloor \frac{\left\lceil \sum_{i=1,\ldots,t} \frac{f_i r_i}{g} \right\rceil + k}{d} \right\rfloor.$$

We have just proved that for any non negative integers $x_1, \ldots, x_t$, if $x_i \equiv r_i \mod gd$, then

$$\phi(x_1, \ldots, x_t) = p_{(r_1,\ldots,r_t)}(x_1, \ldots, x_t).$$

Therefore $\phi(x_1, \ldots, x_t)$ is a quasi-polynomial. $\square$